# A Re-interpretation of Historical References to the Supernova of 1054 AD

by
George W. Collins, II, William P. Claspy, and John C. Martin
Department of Astronomy Case Western Reserve University 10900 Euclid Ave.
Cleveland OH 44106-7215 Collins@Grendel.astr.cwru.edu

**Abstract.** In this paper we re-examine historical references to the supernova event of 1054 AD with a view to establishing a sequence astronomical events which minimizes apparent conflicts between the various sources. We find that the explosion of the supernova is likely to have occurred weeks to months before the commonly accepted date of July 4$^{th}$ 1054. This view is strongly supported by a number of European references to events visible in the evening sky during the spring which are likely to be associated with the appearance of the supernova. We find the best fit to the light curve based on Chinese observations and a maximum visible apparent magnitude for a supernova located at the distance to the Crab Nebula also confirms the earlier explosion date.

**Introduction**

   Ever since the suggestion by Hubble (1928) and Mayall (1939) that the Chinese Guest Star of 1054 AD be interpreted as a Supernova associated with the Crab Nebula, there have been exhaustive searches to find additional corroborating references in the world literature. Efforts by Ho et al. (1972), expanding on the earlier work of Duyvendak (1942), recount references in both the Chinese and Japanese literature. Brecher, et al. (1978) describe a near east account of an Iraqi physician, Ibn Butlān as recorded by Abī Uṣaybiᶜa in about 1242 AD, connecting the event to a plague in Egypt and Constantinople in 1054-5 AD, while Brandt and Miller (1979) make an excellent case for the stellar explosion being seen and recorded by Native North Americans. The absence of any reference in the European literature has traditionally been something of a mystery. Records of the Supernova of 1006 AD and the appearance of Halley's Comet in 1066 AD demonstrate that astronomical phenomena were noticed and recorded by 11$^{th}$ century Europeans. One of the traditional explanations for the absence of European references as expressed by Williams (1981) is that the weather was bad. However, since the supernova would have been remarkably bright and visible for well over a year, that explanation seems rather implausible. Thomas (1979) and separately Zalcman (1979) suggest that the absence is the result of some sort of censorship by the Roman Catholic Church connected with the Eastern Schism generally dated to July 16$^{th}$ 1054 AD. This view is most recently echoed by Lupato (1995), and Giovanni et al (1992) who cite other examples such censorship during the 11$^{th}$ century.

However, we suggest that there are indeed references in the European literature emanating from that time and they serve to add to the information concerning the appearance of the Supernova. Williams (1981) describes an interesting reference in the Rampona Chronicles that was pointed out to him by Newton (1972). Several references dealing with the death of Pope St. Leo IX are discussed Guidoboni et al (1992) and by Breen and McCarthy (1995). McCarthy and Breen (1997) further discuss references in the Irish annals which appear to refer to astronomical phenomena possibly including the Supernova of 1054 AD. Finally there is an Armenian report discussed to by Astapovich (1974) which seems to refer to the Supernova.

The traditional approach concerning reports that might relate to the Supernova has been to decide which reference is the most definitive and likely to be the most accurate. However, all of the references available are derivative from earlier sources and were written in their present form centuries after the event. Therefore it is not surprising that inconsistencies occur between and often within specific reports. We shall review these reports to see if there is a series of astronomical events that minimizes the apparent conflicts between them. In order to understand the sources, it will be necessary to allow for the societal constraints of the time and locale where they were originally recorded. Mindful of the arguments of Guidoboni et al. (1992), we should be careful of the prejudices introduced by the translation of early documents into contemporary language. We will find that some European sources are metaphorical and occasionally anthropomorphic in contrast to the oriental sources, since the astronomical knowledge of the dark ages in Europe was more primitive than in the East. However, we should remember that the level of science in the Orient was far from what we would consider to be science today. A primary motivation for heavenly observation was to search for 'signs' or omens, but incorporated rather 'accurate' positional measurements. Indeed, as pointed out by Ho et al. (1972) the Chinese did not regard the stars as fixed. However, it is likely they would connect an occurrence at dusk with a similar occurrence in the dawn sky at a later date. This is unlikely to be true in medieval Europe.

We begin with the assumption that the supernova did indeed occur in 1054 AD resulting in today's most studied Crab Nebula. This will enable us to incorporate astronomical constraints in developing a sequence of events minimizing the apparent conflicts among the sources. For example we shall see that the Crab Nebula would have been in conjunction with the sun on or about May 27, 1054 AD. Thus, had the Supernova exploded prior to that date, it would have been invisible for a period of time when it was close to the sun. Recent advances in almanacs for small computers make it possible to accurately describe the sky as seen from any location on the Earth a millennium ago. We have used a program known as Redshift I (see Maris 1993) to reproduce the sky as seen from various locales during the eleventh century. Work by Upgren (1991) and Schaefer (1993) make it possible to estimate the effect of the dawn sky on the appearance of astronomical objects at the time of the reported sightings. We shall begin by reviewing and interpreting a number of European sources which suggest sightings of the Supernova as an event in the evening sky before conjunction with the sun in late May of 1054. We shall then briefly review reports from the East which are generally regarded as definitive for establishing the Crab Nebula supernova as occurring in 1054 AD. It will be apparent that conditions of stellar visibility in the dawn sky during the summer of 1054 cast doubt on the traditional interpretation of the Chinese manuscripts. Finally we will apply astronomical constraints applicable to the Crab supernova in order to establish a sequence of events which minimize apparent conflicts between various reports of the time.

**European References**

Breen and McCarthy (1995) discuss several interesting possible European references to the supernova collected by Guidoboni et al. (1992). They cite three references to the death of pope Leo IX the most relevant of which is from the *Tractus de ecclesia S. Petri Aldenburgensi* (see Holder-Egger 1898). The third reference to *De obitu Leonis* by Libuinus , a sub deacon in the Roman Church (see Watterich 1862 p. 176), describes the soul of Leo being taken by angels up to heaven "as along a path strewn with shining garments and lit by innumerable brilliant lamps". The second reference to *Desiderii abbatis Casinesis...* also cited by Watterich (1862) seems too metaphorical to be clearly connected to the supernova, but is consistent with it. Breen and McCarthy (1995) dismiss these with the largely *ad hominum* argument that one should not prefer "an eleventh century cleric in an exuberantly papalist frame of mind being substantially more accurate in his observation of SN1054 than the Chief of the Astronomical Bureau in China". A more sober evaluation of the state of the Roman Church during the eleventh century as compared to the astrologically motivated Chinese observation might call for a less extreme comparison.

It should be noted that at the time of the death of Pope Leo IX (April $19^{th}$ 1054) there was an extensive display of planets in the western sky. Jupiter, Venus, Mars and Mercury were all visible at the same time along with the bright stars of Orion. It is not unreasonable that these could well be *the innumerable brilliant lamps* referred to in the third reference. Should the supernova have appeared, it would have only contributed to the show. The first of these references refers to "an orb of extraordinary brilliance" which briefly appeared at the "very hour" of the death of Pope Leo IX. Watterich (1862) makes it clear that the canonization of Pope Leo IX proceeded with considerable alacrity. A comparison of his accomplishments with those of his predecessors makes it clear why this should have been so and some of the "exuberance" referred to by Breen and McCarthy (1995) may perhaps have appeared in the form of 'poetic license' regarding the precise timing of the event. However, the description of the 'miracles' required for sainthood would have been founded in widely accepted fact.

Perhaps the most unambiguous European reference to the supernova can be found in the *Rampona Chronicle*. As stated by the general editor, Sorbelli (1905), the *Corpus Chronicorum Bononiesium* is a comparison of two $14^{th}$ century compilations, the Rampona Chronicle and the Varignana Chronicle. The former is a history from the beginning of the world to the present (1425) while the latter, written in the Italian vernacular, covers a shorter period of time. As mentioned by Newton (1972) and discussed by Williams (1981), the Rampona Chronicle describes a bright star appearing about the time of the supernova. The Rampona Chronicle as complied by Muratori in the late $15^{th}$ century (see Sorbelli 1905) is derivative from a variety of sources which now appear to be lost. The relevant Latin passage quoted below with our translation clearly has trouble with dates as exhibited by the curious mixture of Roman and Arabic numbers which Williams (1981) cites as common in the 14th and 15th centuries.

> *Anno Christi Ml8 Henricus tertius imperavit annis xl9. Hic primo venit Romam in mense maii.*
> In the year of Christ 1058 (1055) emperor Henry III was in his 49th (39th) year. He initially came to Rome in the month of May.
> *Cuius tempore fames et mortalitas fuit fere in universa terra.*
> At this time there was famine and death throughout the entire land.
> *Et obscedit civitatem Tiburtinam diebus 3 mense iunii.*
> He occupied the Tiburtinam State (He stayed in the town of Tivoli, east of Rome) for three days in the month of June.
> *Hic Henricus pater fuit matris comitisse Mathilde, ex qua Bonifacius marchio genuit ipsam Matheldam. Tempore ipsius Henrici.*
> This Henry was protector of the mother of the countess Mathilda (Beatrice, widow of the duke of Tuscany and sister of Henry III) from whom the Marquis Boniface begot Mathilda herself. This was the time of Henry.
> *Tempore huius stella clarissima in circuitu prime lune ingressa est, 13 Kalendas in nocte inito.*
> During this time (Henry's) an extremely bright star appeared in the circuit of the new moon (at the location of the new moon, i.e. east of, and close to the sun) on the beginning of the night of the 13th of the kalends.
> *Huius tempore Hildebrandus cardinalis, qui postea Gregorius papa factus est, cum legatus esset in Gallia et in concilio contra multos simoniacos episcopos processisset,…*
> At this time Cardinal Hildebrand who later became Pope Gregory was in council as papal legate in France where he moved against those who were Bishops by way of simony,…

We have placed in parentheses corrections or explanations of certain aspects of the text. The chronicle continues from this point with a detailed description of Hildebrand's actions against simoniac bishops in France. It is reasonably clear from the description that these were actions taken under the auspices of Pope Victor II in 1055 AD specifically at the councils of Lyons and Chalons where Hildebrand with full legatine authority dealt with several simoniac Bishops. An alternative could be the Council of Sens which concerned the heresies of Berengarius and his philosophy concerning the celebration of the Holy Eucharist. This Council was also presided over by Hildebrand with full legatine authority in 1054 AD.

The opening reference with the dates of ml8 and xl9 probably refer to reading mliiiii as mlviii and possibly xxlix as xlix since the former dates correspond to the actual numbers appropriate for the text. Emperor Henry III did indeed travel to Italy in the spring of 1055AD for the installation of Pope Victor II in Florence among other things (see Steindorff 1881). After the death of the Boniface, Marquis of Tuscany, in 1053, Henry took on the role of "father protector" for his sister the widow Countess Beatrice. Hughs (1948) notes that Henry found it necessary in 1055 to dissolve her subsequent secret marriage to Duke

Godfrey II of Lorraine who had been at odds with the Emperor for some time. This may have been because Tuscany was the center of opposition to the reform of the Roman Church instigated by the German Popes appointed by Henry and he simply could not afford a union between Tuscany and powerful Lorraine. This would account for the references to the marriage in a source dealing largely with the affairs of Henry III.

It is interesting that famine and plague are also mentioned as occurring at this time in the Near East reference discovered by Brecher, et al. (1978). The general temporal reference to a bright star near the location of the new moon is in keeping with the chronology of this section. However, since the reference clearly refers to the beginning of the night, it is likely that the date is *13 Kalendas Juni* or May 20$^{th}$. As described above the supernova would have been 7 degrees east of the sun at a location commonly associated with the new moon.

The diversity of sources noted by Sorbelli (1905) is demonstrated by the shift in text and style after the passage concerning the star. This suggests a change of source from a secular one documenting the affairs of Henry III to a clerical source concerned with the detailed actions of the church. However, it is interesting that no mention here is made of the major events in the church of 1054 - notably the death of Pope St. Leo IX on April, 19$^{th}$, 1054 AD or the Eastern Schism usually dated to July 16$^{th}$ 1054 AD. This suggests that the entry point to the now-lost clerical source is likely to be 1055 AD. While it is tempting to dismiss a chronicle complied so far from the event and containing numerous errors of dates, it must be remembered that the Chinese and Japanese sources are also derivative sources complied at a much later date than the event.

To understand the significance of the Rampona Chronicle, it is necessary to look at the material surrounding the reference to the star. Virtually all the references before and after the section dealing with Henry III and the star are clerical. Clearly the references to the star and the concerns of the Emperor Henry III were inserted into a church history from a different source. The phrase *Tempore ipsius Henrici* suggests that specific dates were relatively unimportant to the original author. However, the choice of the Latin *stella clarissima* to describe the star is as superlative as can be expressed suggesting that this was indeed a very bright star.

Breen and McCarthy (1995) translate the words *prime lune* as refering to the "first day of the calendar or ecclesiastical moon…". Since there are no new moons on the 13$^{th}$ of the kalends they reject the entire reference out of hand. They further note that the date disagrees with the Chinese date of July 4$^{th}$. This interpretation clearly avoids the problem which led Williams (1981) to state that he could make no sense of the phrase *in circuitu prime lune ingressa est*. As appears clear from the style, a secular account would be less likely to refer to the ecclesiastical calendar, than a cleric. Our interpretation that the phrase denotes a place in the sky rather than a calendar date avoids this problem and makes this a reference to a dusk sighting most likely on May 20$^{th}$ when the supernova would have been just where one would look for the new moon. Since the events described in the chronicle match actual historical events, it seems plausible to suggest that this indeed represents a European sighting of the Supernova of 1054 AD.

Finally, there are two references from eastern Europe which are important. The first, discussed at length by Brecher et al. (1978), describes the report of an Iraqi physician Ibn Butlān as recounted two centuries later by Abī Uşaybiᶜa. Two references to the end of the Hegira year of 445 and 446 corresponding to (April 23,1053- April 11, 1054) and (April 12, 1054 - April 1 1055) are given in different sections of the recounting by Abī Uşaybiᶜa. The latter range is generally preferred since it agrees with the Chinese date and subsequent visibility of the event while the former is considered a copying error. However, Guidoboni et al. (1992) take these dates literally suggesting that the earliest date for the supernova should be April 11$^{th}$ 1054AD.

The second reference is to Armenian Chronicles of the IX through XVII centuries complied by Matenadaran [specifically, see the chronicle of Этум Патмич (E. Patmich) in Akopan (1956)]. Astapovich (1974) corrects the dates and interpretation of an earlier translation by Astapovich and Tumanian (1969) where they felt the reports were of bright meteors. He now believes a proper translation should read " .. *1054 of the New Era was the fifth year of the reign of Leo IXth. That year on the moon's disk a star has appeared. It happened on the 14$^{th}$ of May in the first part of the night*". He believes the reported location, time and description more correctly describe the supernova event of 1054 AD. Calculations with the program Red Shift show that the moon and the Crab Nebula were in conjunction on May 11$^{th}$ at approximately 9h UT. This suggests that it is unlikely that the proximity of the moon to the Supernova resulted in the phrase "…*on the moon's disk a star has appeared.*" The words "*That year….*" Also suggest it was more than an ephemeral event such as a meteoroid impact. Astapovich and Tumanian (1969) clearly had difficulty with the translation and interpretation of this passage, but make it clear that the event was very striking and probably represented a new star. An interpretation consistent with their writing is that the star was first noticed on the night of May 14$^{th}$ as a result of either increasing brightness or weather and the phrase "*moon's disk*" may be translated from 'moon's orb', 'moon's orbit', or 'circuit of the moon' indicating a place in the sky such as is the case in the Rampona Chronicles. Unfortunately support for this contention is unlikely to be found without a re-analysis of the original of the Chronicle which is unavailable to us.

**Eastern References**

The primary references to the Supernova of 1054 AD are found in the Chinese and Japanese chronicles. They were comprehensively translated by Duyvendak (1942), re-discussed and added to by Ho et al. (1972), and re-interpreted by Breen and McCarthy (1995). Ho et al. (1972) correctly point out that there are a number of internal inconsistencies in many of these sources while Breen and McCarthy (1995) go to some lengths to establish that many of the sources are derivative from a single earlier source. Duyvendak (1942) and Ho et al. (1972) suggest that the definitive chronicles are the *Sung-shih* and the *Wen-hsien T'ung-k'ao* complied in 1345, and 1280 respectively. Breen and McCarthy (1995) take the *Sung hui-yao* complied by Chang-Te-haiang to be the definitive text even though Ho et al. (1972) give the earliest known text as *Hsü Tzu-chih t'ung-chein ch'ang-pien* whose author Li Tao died in 1184 AD. The suggestion is that many of the later texts derive from this one. As they point out, the date for the event in this earliest text is internally inconsistent and could either be July 4th 1054 AD, June, or August. However, it must be remembered that all the reports cited in the literature were complied after the event and therefore represent a summary of observations of the supernova made well after the explosion.

While the 4th of July 1054 AD has become generally accepted as the date of the explosion of the supernova, as pointed out by Ho et al. (1972) certain anomalies exist in the Chinese references that are regarded as definitive. For example, the position of the 'guest star' is given as "several inches south east of ς Tauri". However, the actual location of the Crab Nebula is about a degree north west of ς Tauri. Ho et al. (1972) make a convincing case for the named reference star actually being ς Tauri. However, the direction to the Crab Nebula is opposite from that given in the chronicles and the angular distance would correspond to much less than "several inches" if Ho et al's (1972) interpretation is correct. On the other hand, if the reference star is β Tauri, the location of the Crab nebula some 6 degrees 45 minutes south east matches the direction and magnitude given by the Chinese record remarkably well. To evaluate the plausibility of the account in referring to ς Tauri, we used Redshift to reproduce the eastern sky for the morning of July 4th 1054 AD. At the beginning of Astronomical Dawn ς Tauri is a mere 3 degrees above the horizon. Upgren (1961) gives a minimum altitude of 6-8 degrees for a third magnitude star to be visible. This result is in full agreement with a more extensive extinction calculation based on the work of Schaefer (1993) described below. We suggest that a star which is too low to be visible at the beginning of Astronomical Dawn is likely not to be seen due to the rapidly brightening sky near the horizon. We will now attempt to quantify this contention.

To definitively describe the visibility of astronomical objects in the dawn sky is extremely difficult as the various obscuring phenomena described by Minnaert (1954), and Schaefer (1993) are difficult to quantify and are extremely complex. However, we may attempt a conservative estimate of the visibility of stars rising in the dawn through the following argument: Schaefer (1993) provides various values for the zenith sky brightness when the sun is below the horizon. In addition, he provides a value for the sky brightness near the horizon at the beginning of Astronomical Dawn. We assume that the eastern sky will brighten at least as fast as the zenith, so that one may project a lower limit for the sky brightness near the eastern horizon during the dawn. That this provides a true lower limit only requires that the zenith brighten no faster than the near-horizon sky. This lower limit is shown in Figure 1. Using Minneart's (1954) estimate for the first appearance of first and fifth magnitude stars as a function of solar altitude below the horizon, we may interpolate the sky brightness that would just mask the appearance of any star of known brightness at the Zenith. In Figure 1 we show these results for both β and ς Tauri. However, on July 4th they will both be near the horizon so that we can expect a higher value of the sky surface brightness to mask their appearance due to the atmospheric extinction of the stars. This can be calculated from Schaefer (1993) by means of equations 3a,b and equation 51 given the zenith distance of the two stars for a particular solar altitude below the horizon. We give the results of this calculation for both β and ς Tauri as seen from Beijing in Figure 1. Since observations made from the more southern location at K'Ai-Feng (e.g. Breen and McCarthy, 1995) will favor the visibility of ς Tauri, we have repeated the calculation for this case. The two visibility curves for ς Tauri are labeled N and S respectively while the visibility curve for β Tauri as seen from Beijing is labeled N. The altitudes for the three cases are indicated by values associated with arrows showing their position on the respective visibility curves for the beginning of astronomical dawn. It is clear that at no time will ς Tauri be visible in the dawn sky from either location in China on the $4^{th}$ of July. However, by virtue of its greater brightness and higher altitude, β Tauri would be visible from either location until nearly nautical dawn (i.e. the solar altitude is -12 degrees). The internal consistency of this procedure is nicely demonstrated by the fact that the visibility curve for ς Tauri crosses Schaefer's dark sky horizon line at an altitude value consistent with Upgren's (1991) empirical value for altitude for minimum dark sky visibility of $3^{rd}$ magnitude stars.

The response to diminishing atmospheric extinction with the rising of the star supports the claim that a star not visible at the beginning of Astronomical Dawn will generally remain invisible to the un-aided eye. Allowing for about a degree of motion along the steeply inclined ecliptic per day, we can use Figure 1 to show that it will be more than a week after the $4^{th}$ of July before ς Tauri becomes visible in the dawn sky. Thus the Chinese reference to ς Tauri as the companion star to the Supernova is either incorrect or it is based on later observations. In that regard, it is worth noting that the chief astrologer did not make his prognostications concerning the "guest star" until August $27^{th}$ (see Ho et al. 1972) by which time the Supernova was no longer visible in the day time and would have faded to a more stellar-like appearance. In addition, it would have been well away from the sun and surrounded by a reasonably rich and presumably familiar stellar field from which to establish its astrological significance. Thus, it would appear that July $4^{th}$ is just one of many observations which went into the Chinese record and it is perhaps presumptuous to attribute any great significance to the date as representing the date of the explosion.

Ho et al. (1972) note that the Japanese references all give the fourth month as the first appearance of the Supernova. One of the references specifically suggests May 20 through May 29 for a date of the appearance of the star. However, Breen and McCarthy (1995) present a protracted argument suggesting that all the Japanese references derive from the same source and there is an error in the month and the date should be June $28^{th}$ -July $7^{th}$ in order to agree with the Chinese date. They rule out the actual date of the fourth month by citing Mayall and Oort (1942) that ς Tauri was in conjunction on the $27^{th}$ of May so that no 'guest star' would be visible. However, the actual Japanese date would place it near Orion in the time interval of May $20^{th}$ -$29^{th}$ .

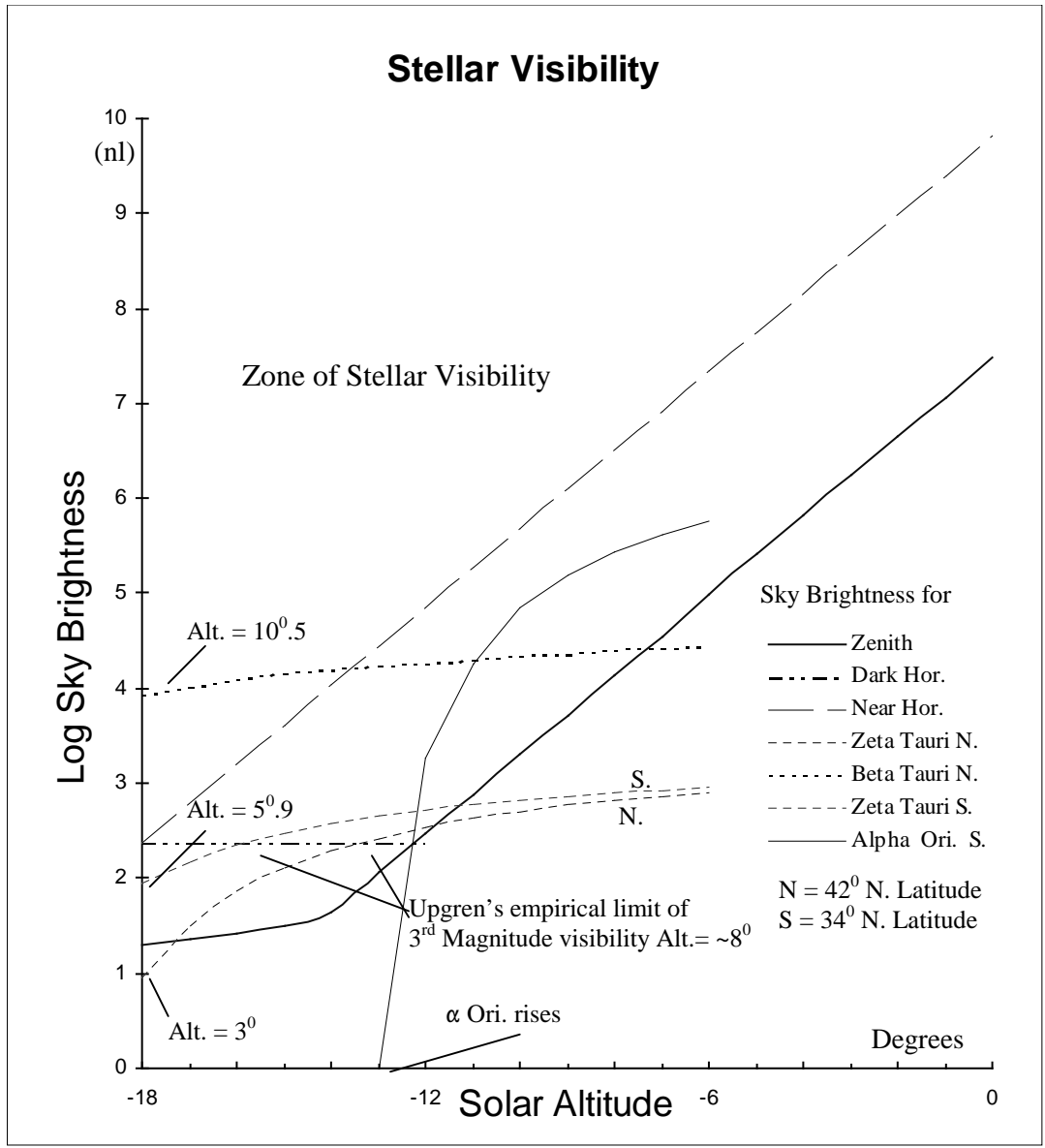

Figure 1 depicts the visibility of β and ς Tauri as well as α Orionis as seen from north and central China during the dawn of July 4$^{th}$ 1054. Only the visibility of ς Tauri is show for both locations. For a star to be visible it must lie above the dashed line representing the sky surface brightness during the dawn. On the 4$^{th}$ of July 1054 A.D. only β Tauri will be visible from either location and even then only before the beginning of Nautical Dawn. The supernova would have been in the zone of visibility well above the figure.

On May 20$^{th}$, the supernova would have been seven degrees east of the sun and would be quite visible as a bright star, near the constellation of Orion as described, at dusk. However, the Japanese reference refers to the star as being seen in the east and as Breen and McCarthy (1995) point out Orion is not visible in the morning sky during June and most of July. We confirm this by calculating the visibility of α Orionis on the fourth of July and displaying the results in Figure 1. It is clear that this brightest Orion star would not be visible in the morning sky until late July or August. This apparent inconsistency in the Japanese references can be understood by consulting the translation of the *Mei Getsuki* by Xi and Bo (1966b) which we repeat below:

> *After the 2$^{nd}$ third of the 4$^{th}$ month, the second year in the Ten Ki period of Japan, at the time of Chhou, a guest star appeared three times at the Hsiu Tsui (Turtle). It was seen in the east, "with Ten Kwan Hsing, as big as Juipter."*

The references to multiple appearances suggests that this report spans some time. This could include evening observations during the stated fourth month when it would have been seen in the vicinity of Orion during the 20$^{th}$ -29$^{th}$ of May as well as a later time when it appeared in the eastern sky at dawn near *Ten Kwan Hsing*. These authors make this aspect of the translation a separate sentence whereas the translation presented by Ho et al. (1972) which is repeated by Breen and McCarthy (1995) and both of which derive from Duyvendak (1942) include this as part of the previous sentence and the reference to "three times" is absent. The presence of the "three times" also indicates that the record is a summary of observations made well after the event.

Finally there are two Chinese references which relate the appearance of a 'guest star' to a solar eclipse in 1054. The first of these is the *Ch'I-tan-kuo chih* (Text D of Breen and McCarthy 1995 ) written during the middle of the 13$^{th}$ century. The second is the *Sung-shih hsin-pien* prepared by K'o Wei-ch'i in the 15$^{th}$ century. Ho et al. (1972) give a translation of the latter as:

> *"During the first year of the Chih-ho reign-period [1054] there was a solar eclipse at midday and a guest star appeared within the Mao [lunar mansion]:[the Pleiades]"*.

They further explicitly state that no date is given other than the year and that the passage is the same as that given in the first reference stated above.
Their translation of the former is:

> *"During the eighth month [of the twenty-third year of the Chung-his reign-period][1055] the King passed away... Previously there was a solar eclipse at midday and a guest star appeared within the Mao [lunar mansion][the pleiades]. The Assistant Officer in the Bureau of Historiography, Liu I-shou, said, 'Isn't this an omen that [the King of Ch'I-tan] Hsing-tsung will die?' The prediction did come true. The same passage is given in Liao-shih-i"*.

Breen and McCarthy (1995) give a translation as:

> *"(In the 23$^{rd}$ year of the period Ch'ung-his) in the 8$^{th}$ moon the lord of the country died.... Previously there had been a sun-eclipse, and in the 1$^{st}$ moon (January 31$^{st}$- Feb 28$^{th}$ 1055) a guest-star had appeared in the Pleiades. Liu Yi-sou, Senior Vice President of the Bureau of Historiography, said "Now Hsing-tsung has died, (these omens) have indeed come true!"*

Ho et al.'s (1972) translation suggests that the sighting of the guest star occurred during the solar eclipse. Breen and McCarthy's (1995) version suggests that the guest star sighting had nothing to do with the solar eclipse and occurred in January 1055. If the Ho et al.'s (1972) translation is correct, then a firm date and location for the sighting can be found from the only solar eclipse in China in 1054 which occurred on May 10$^{th}$. Had the Supernova exploded in April or early May, it certainly would have been visible during totality. We now consider the extent to which astronomical information regarding supernovae may relate to the various references of the visibility of the supernova during the Spring and Summer of 1054AD.

**Astronomical constraints**

Much has been discovered about supernovae since the early discussions of Mayall and Oort (1942). Determination of the expansion age of the Crab Nebula consistently give origin dates in the 12$^{th}$ century (e.g. Trimble 1968, Wycoff and Murry 1971, and Nugent 1998 ) and has been used by Williams (1981) and others to suggest that the event in 1054 AD should not be associated with the Crab Nebula Supernova. However, contemporary thought suggests that the expansion age is a lower limit due to secular acceleration of the nebula (see Trimble and Woltjer 1971 and Bientenholz et al. 1991). Other properties of the Crab Nebula such as the distance (e.g. Trimble 1968, 1973) and reddening (e.g. Miller 1973) have been determined so that we may assess the likely maximum apparent magnitude of the various types of supernovae appropriate for the supernova that led to the formation of the Crab Nebula. The light curves appropriate for various types of supernovae may then be compared to observations from the time. We have the following points on such a curve. Ho et al. (1972) conclude from *Sung-shih* and the *Wen-shien t'ung-k'ao* that the duration of visibility was from July 4$^{th}$ 1054 AD to April 17$^{th}$ 1056 AD. The *Sung-hui-yao* suggests that the duration of daytime visibility was 23 days. Taking the initial observational date as July 4$^{th}$, the Supernova would have been about 60 degrees west of the sun 23 days later so that a comparison with Venus at Greatest Elongation, and which is visible in the day time, is not inappropriate. Thus we may expect the Supernova to have been about -3.5 ± 0.5 $m_v$ at the end of July when it disappeared from the daytime sky. That suggests a drop of about 10 magnitudes in the 630 days when it disappeared from the night sky on April 17$^{th}$ 1056 AD. However, other Chinese reports only specify the time of disappearance to the third month so that a duration of 615 ± 15d seems a prudent range for the interval between the disappearance from the day- and nighttime skies. We will take the visual magnitude at disappearance to be 6.0 ± 0.5 $m_v$. Unfortunately the acceptable range in distances (e.g. Trimble 1973) introduces a range of half a magnitude in the apparent magnitude at maximum. So combining the maximum absolute magnitude for supernova with a distance of 2 ± 0.5 kpc for the Crab Nebula, one would expect an apparent visual magnitude at maximum to be between -6.0 and -8.5. Correcting for interstellar extinction from Miller (1973), these values are reduced to -4.5 to -7.0 visual magnitude nicely bracketing the traditional value of -5 used by Mayall and Oort (1942). Even these reduced values would make the supernova the brightest object in the sky other than the sun or moon that anyone of that time had ever seen. The twinkling of a point source of such brightness would indeed appear to have "pointed rays shot out from it on all sides" as described by the Chinese. However, it is highly likely that if maximum occurred near conjunction with the sun, it would not be seen for several weeks. In addition, to fit theoretical light curves we will require the magnitude at maximum to be that appropriate for the type of Supernova located at the distance (e.g Trimble 1973) of the Crab Nebula and reddened according to Miller (1973).

In Figure 2 we show the best fit between V- light-curves for typical supernovae of various types and the estimated brightness for the Chinese dates. The light curves were obtained from Doggett and Branch (1985). Only by taking the intrinsically faintest type supernova at the greatest distance can we arrange to have the maximum near July $4^{th}$. The typical rapid decline to the logarithmic decay curve for all but those supernovae exhibiting a high level plateau would force the maximum to be well before July $4^{th}$. The presence of a plateau such as the Type IIP would push the date of maximum to be even earlier. Therefore, it seems clear that the maximum must have occurred before July $4^{th}$ for a drop to about -3.5 $m_v$ from the maximum of any type supernova in substantially less than 23 days is unlikely. There can be little doubt that the Chinese observers whose competence is so vigorously extolled by Breen and McCarthy (1995) and who could follow a bright star rising into daylight from the darker dawn sky with astrolabes (see Ho et al. 1972) would be able to see a daytime object of -3.5 $m_v$ if not fainter. Thus the range of ± 0.5 $m_v$ is entirely reasonable for this point. The competence of the Chinese observers also suggests that the visual error limits for the disappearance from the nighttime sky are also reasonable.

The strongest constraint on the light curve is placed by the two points marking the disappearance from the day and night time skies. It is difficult, but not impossible to fit this slope with typical Type II light curves. Even Type Ib fades two quickly to readily fit the curve. However, it should be noted, as Figure 2b shows, the light-curve for Type Ia fits remarkably well with the most probable peak occurring around May $14^{th}$. The effect of the conjunction with the sun for this probable peak is shown in the inset of Figure 2b utilizing the visibility curves of Shaefer (1993) for bright objects close to the sun.

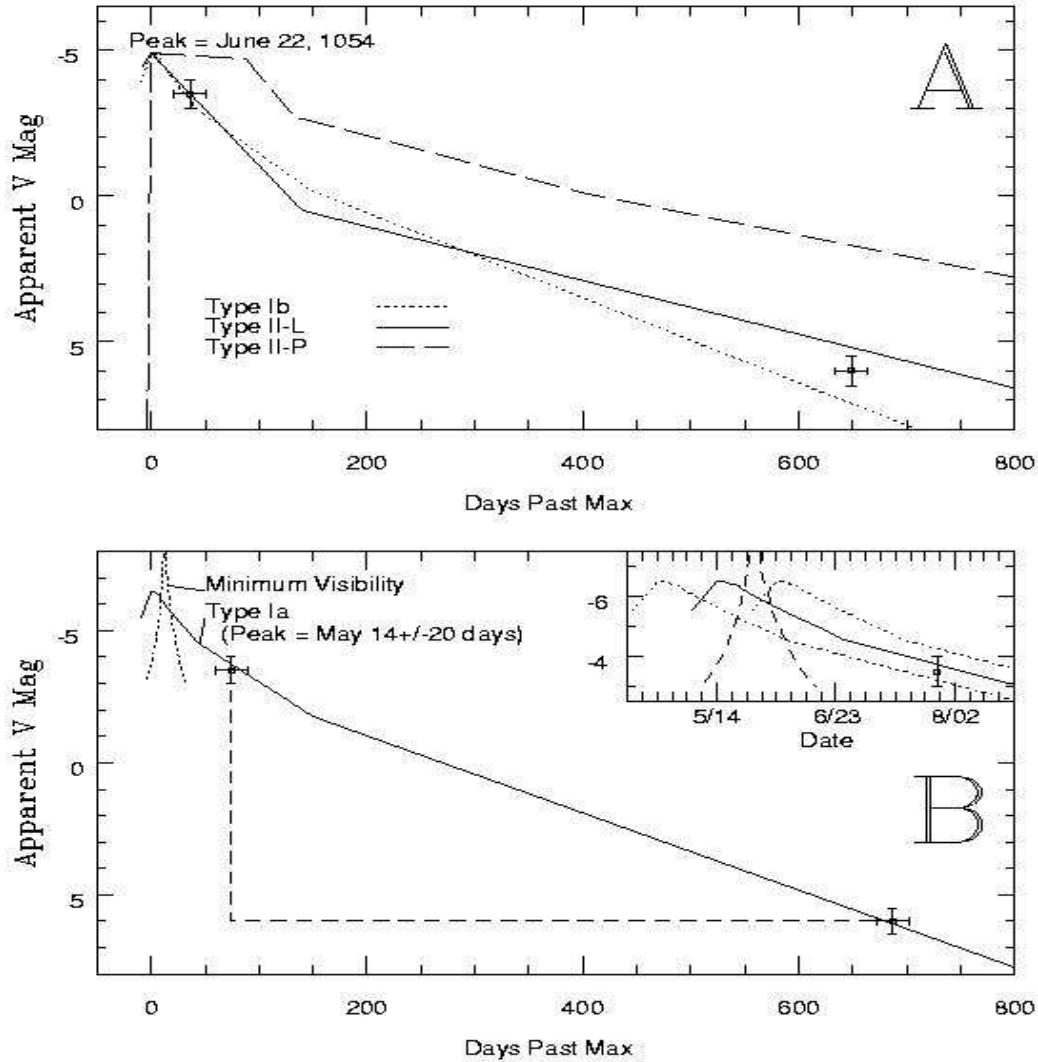

Figure 2 shows a comparison between light curves for various types of supernovae and the apparent brightness of the supernova of 1054AD. In panel **A** we show that Types IIP, L and Ib fail to fit the apparent brightness appropriate for the disappearance from the daytime and night skies at the dates given by the Chinese reports. The fits are also constrained by the appropriate maximum apparent magnitude for these supernovae located at the distance of the Crab Nebula. Panel **B** shows the same comparison for a supernova of Type Ia. In addition, we have shown the minimum magnitude for daytime visibility for objects near the sun as given by Schaefer (1993) indicating that daytime sightings would have been unlikely near maximum brightness due to the near conjunction with the sun. The insert in panel **B** shows the likely light curve for the supernova vs calendar date which is consistent with the astrophysical and observational constraints. We have taken the temporal error to be about 20d from the temporal uncertainties of the observations combined with reasonable uncertainties for the minimum magnitude of visibility in the day and night time skies.

While some have suggested that the large temporal interval for the two Chinese points is greater than any supplied by contemporary observations, it should be noted that Kirshner and Oke, (1975) followed Type Ia supernova 1972e for over 700 days and found no departure from the linear decline in brightness. Minkowski (1964), Shen (1969) and others have also noted that Type Ia light curve fits the Chinese points better than any other options. However, Minkowski (1968) found other reasons to question the Crab progenitor being Type Ia and ends his review article with the statement *"The available data are too scanty to permit the assignment of a type to the supernova of +1054."* Certainly contemporary models for Type Ia supernovae leave no residual pulsar and therefore make it difficult to declare the Crab to have been a Type Ia supernova. Nevertheless, the remarkable agreement of the light curves with the Chinese points must be added to the growing list of anomalies for the Crab. Apparently abundances cannot definitively assign a supernova type to the Crab (Luck 1998). Van den Berg (1973) notes that the expansion velocities agree better with Type Ia expansion velocities rather than Type II and Zimmer (1998) has reviewed the well known 'missing mass' discrepancy suggesting that the mass of the Crab is only a few percent of that expected from a Type II supernova explosion. Perhaps it is possible that these anomalies can be resolved through the extreme variability of Type II supernovae or other variations of Type I so we adopt Minkowski's (1968) view given above. However, since noting the accuracy of the fit of the Type Ia light curve to the astronomical constraints is not the same as assigning the actual type, we shall proceed to use that curve to provide an estimate for the brightness throughout the duration of the event.

A maximum apparent brightness of about -6 would have made this event extremely spectacular. However, the conjunction of the supernova with the sun on May $27^{th}$ means that even a maximum visual magnitude of -6.5 would have been obscured by the sun for some time. Using Schaefer's (1993) estimate of daytime visibility of objects near the sun, we show the minimum magnitude for day time visibility of the supernova during the time of conjunction relative to the optimum date for maximum brightness of the supernova. The sub-panel in Figure 2B shows the period near maximum with the most probably calendar date for the occurrence of maximum brightness. The temporal errors are indicated by the dotted light curves either side of the most probably fit. As in the main figure, the dashed line is the minimum magnitude for the visibility of objects near the sun during the time of conjunction from Schaefer (1993).

We can now suggest the following sequence of events which minimizes conflict between and within the historical references in a manner that is consistent with known astronomical constraints. In April or early May of 1054 AD the supernova which has given rise to the Crab Nebula exploded. The rise time to maximum brightness for a Type I supernova is poorly known but is expected to be about several weeks while that for Type II is even longer ranging to months. If one takes the error limits to their extreme, it is possible that the supernova exploded in early April which would make Brecher et al.'s (1978) correction to the date given by the Iraqi physician Ibn Butlān, unnecessary for it would have just appeared at the end of the Hegira year of 445. This appears to be the view of Guidoboni et al. (1992) who take April $11^{th}$ as the first sighting of the supernova. A somewhat later explosion would still have been early enough for the supernova to have provided a heavenly sign for the canonization of Pope St. Leo IX with only minimal poetic license. It even allows for the date of the feast of St. George (April 24$^{th}$) given in the Irish annals of the extremely metaphorical report discussed by McCarthy and Breen (1997) to be accommodated. An early explosion date allows for the supernova to be visible during totality of the solar eclipse of May 10$^{th}$ and shine brightly in the dusk during the last three weeks of May. The Armenian report in the evening sky on the 14$^{th}$ of May is then easily accommodated as are the Japanese reports of a new star appearing in the vicinity of Orion during late May. It also fits the Rampona Chronicle perfectly for the supernova would have been just 7 degrees east of the sun on the 20$^{th}$ of May in the very location where one looks for the new moon. Maximum brightness is liable to have occurred near conjunction with the sun thereby minimizing its impact on observers. The Japanese and Chinese saw it during late June and July. The identification by the Chinese on July 4$^{th}$ is likely to have been made with respect to β Tauri which would have been about the only star visible close to the Supernova in the morning sky. The reported position is in remarkable agreement with the location of β Tauri. Later, the proximity to ζ Tauri would have it replace β Tauri as the identifier. Much later (August 27$^{th}$), when the stellar field was more complete, the astrologers would be confident enough to interpret the event for the emperor. By then the star would have no longer been visible in the daytime sky and would properly be resembling a "guest star" rising before dawn. For astrological reasons, the Chinese would have continued to observe the guest star until it disappeared from the dark night sky a year an a half later. For convenience we summarize this sequence in Table 1.

Table 1
A Chronology of possible Supernova Sightings

| Date | Location | Source | Appearance |
|---|---|---|---|
| April 11$^{th}$ | Constantinople | Diary of Ibn Butlān | Star |
| April 19$^{th}$ | Flanders | *Tractatus de ecclesia* | Bright Light |
| April 24 | Ireland | Irish Annals | Fiery Pillar |
| Late April | Rome | *De obitu santi Leonis pp IX* | Bright Light |
| May 10$^{th}$ | China | *Sung-shih hsin-pien* | Star |
| May 14$^{th}$ | Armenia | Этум Патмич | Star |
| May 20$^{th}$ | Italy | Rampona Chronicles | Very Bright Star |
| Late May | Japan | *Mei Getsuki* | New Star |
| June | Japan | *Mei Getsuki* | Star |
| July 4$^{th}$ | China | *Sung hui-yao* | Star/Comet |
| July 27$^{th}$ | China | *Sung hui-yao* | Star |
| April 1055 | Constantinople | Diary of Ibn Butlān | Star |
| April 17$^{th}$ 1056 | China | *Sung hui-yao* | Star |

It is worth noting that during the early days of the supernova a point source of around -5 m$_v$ or brighter would be an amazing sight. Near dawn the twinkling and color variation due to atmospheric refraction would have made the object appear even more active. It would not be surprising if it were interpreted as showing a disk which may account for the use of the word "po" usually reserved for comets by the Chinese. Later, as it dimmed, it would become a "guest star". Finally, after nearly two years, the star would no longer have been visible in the night sky leading to the final entry from the Chinese reports.

**Conclusions**

We have reviewed reports from the 11$^{th}$ century regarding events which can be associated with the explosion of a supernova during 1054 AD the remnants of which are most certainly what is known as the Crab Nebula. The interpretation of the section of the Rampona Chronicle dealing with the appearance of a very bright star is new and provides a level of internal consistency strengthening the view that this is an evening sighting of the supernova. We have offered a sequence of astronomical events consistent with most aspects of the various reports including a number from Europe. These reports when combined with the oriental reports strongly suggest an explosion date of mid-spring 1054 AD. However, all of the European reports refer to events in the evening sky during the Spring of 1054 AD while none found so far refer to events in the morning sky of mid June and beyond.

This curious absence still leaves us with a bit of a mystery. Indeed, since the references to earlier evening phenomena clearly demonstrate that such heavenly events were of interest to the residents of medieval Europe, the mystery surrounding the lack of reports describing the later and more pervasive morning-event is only increased. This state of affairs would seem to lend credence to the "conspiracy view" suggested by Thomas (1979), independently elaborated on by Zalcman (1979), and further developed by Lupato (1995) that such writings were suppressed by the Roman Catholic Church. The formal separation of the Church of Rome from the Eastern Orthodox Church, known as the Great Schism, is usually dated from the excommunication of the Eastern Patriarch Michael Cerularius, Emperor Constantine Monomachus and their followers on the 16$^{th}$ of July 1054 AD by three legates from the Church of Rome. This event does coincide with the supernova being brilliantly visible in the dawn sky and on into the day. According to Runciman (1955), the Schism, while of minor importance in the East, was viewed as a most important event in the West so that events surrounding it deserve some attention.

While the stated mission, which brought the three Roman Legates to Constantinople in the summer of 1054 AD, was to form an alliance between the Church of Rome and the Eastern Church by smoothing over past differences (e.g. Runciman 1955, or Steindorff 1881), an excellent case can be made for the two central parties to the Schism wanting the opposite. Cardinal Humbert, who headed up the legation to Constantinople, made requirements which he knew the Eastern Patriarch would not accept. On the other hand The Eastern Patriarch, Michael Cerularius, regarded the Roman See as morally bankrupt and controlled by German barbarians. He clearly desired primacy over his own church. For a variety of reasons, including knowledge of the death of Pope Leo IX in April (e.g. Every 1962) and the legate's ties to the duplicitous Marianos Argyros, military governor of the Greek colony in southern Italy (e.g. Runciman 1955 and Gilchrist 1993), he distrusted the legates and doubted the authenticity of their credentials.

Zalcman (1979) points out that Ibn Butlān, the Iraqi physician, Nestorian Christian, and original author of the Constantinople reference to the supernova (e.g. Brecher et. al. 1978) was a confidant of the Eastern Patriarch so there can be little doubt that the rather mystical (see Runniciman 1955) Michael Cerularius was aware of the new star in the sky. It takes very little imagination to see how he might have used such an omen to support his suspicion of the papal legates. The legates, on the other hand, would have preferred that such arguments and their visible proof not be subsequently noted in the West. Thus, perhaps it is not surprising that the only "eyewitness" record to what both sides now seem to regard as a somewhat sordid affair is that of the chief legate and papal secretary Cardinal Humbert de-Silva. Some two years later following the death of Pope Victor II, another legate to Constatinople, Frederick of Lorraine, became Pope Stephan IX (X) with the help of Cardinal Humbert. Pope Stephan IX then elevated Cardinal Humbert to Chancellor of the Roman See and Vatican Librarian. Cardinal Humbert's detailed account of the journey to Constantinople contains no reference to the star.

While this argument is largely circumstantial, it does provide a basis for understanding the lack of subsequent references to the supernova of 1054 AD in the largely clerical European literature. It is also a superior to the "poor weather" hypothesis suggested by Williams (1981) and suggests if further references are to be found they are likely to reside in the secular literature and historical chronicles whose origin is other than the Church of Rome.

The authors would like to acknowledge Dr. Donald Poduska, Fathers Paul Sciarotta and Timothy Plavac, Drs. Gerard Angoustures, and Ann Fry for their help in translating various sections of the Latin and Italian documents. Further thanks are due Dr. Poduska for additional insights into the medieval mind. The authors are also deeply grateful to Christa Luck for helping with the translations of the convoluted, but fascinating, German contained in Steindorff (1881). Final thanks are due Professor Shmaryu Shvartsman for his translation of the Armenian report.